\documentclass[epj]{svjour}
\usepackage{graphics}
\usepackage{microtype}
\usepackage{mathrsfs}
\usepackage{bm}
\usepackage{amsmath}
\usepackage{amssymb}
\usepackage{url}
\usepackage[sort&compress, numbers]{natbib}
\usepackage[colorlinks=true, citecolor=blue,linkcolor=blue]{hyperref}
\usepackage{enumitem}

\usepackage{enumitem}
\setlist[itemize]{itemsep=0.5\baselineskip, parsep=0.5\baselineskip}
\usepackage{microtype} 
\usepackage{verbatim} 

\begin{document}
\title{Model Study of Eigen-Microstate Signatures of Criticality in Relativistic Heavy-Ion Collisions}
\author{Ranran Guo\inst{1}\inst{2} \and Jin Wu\inst{3} \and Mingmei Xu\inst{1}
\thanks{\emph{Email address:} xumm@ccnu.edu.cn}%
\and Zhiming Li\inst{1} \and Zhengning Yin\inst{1} \and Yanhua Zhang\inst{4} \and Jinghua Fu\inst{1} \and Yuanfang Wu\inst{1}\thanks{\emph{Email address:} wuyf@ccnu.edu.cn}
}                     
%
%
\institute{Key Laboratory of Quark and Lepton Physics (MOE) and Institute of Particle Physics, Central China Normal University, Wuhan, 430079, Hubei, China \and College of Information Engineering, Zhengzhou University of Technology, Zhengzhou, 450052, Henan, China \and College of Physics and Electronic Information Engineering, Guilin University of Technology, Guilin, 541004, Guangxi, China \and Department of Physics and Electronic Engineering, Yuncheng University, Yuncheng, 044000, Shanxi, China}
%
%
\abstract{
We present a systematic model study of the eigen-microstate approach (EMA) for characterizing critical-scaling-motivated correlations in relativistic heavy-ion collisions. Using UrQMD and two stochastic reference models, we investigate the influence of conventional non-critical correlations on eigen-microstate patterns and eigenvalue spectra. Multiscale momentum-space correlations are introduced through event-level and particle-level incorporation of Critical Monte Carlo (CMC) samples into the UrQMD background. The CMC component produces characteristic cluster-like eigen-microstate patterns and enhances the weights of leading fluctuation modes, with a stronger response observed for event-level incorporation. These patterns persist across different momentum-space resolutions, while the ratios of leading eigenvalue weights exhibit approximately resolution-independent behavior at sufficiently large CMC fractions. Our results demonstrate the potential of EMA for extracting multiscale momentum-space correlation structures from non-critical heavy-ion collision backgrounds.
\PACS{
      {PACS-key}{discribing text of that key}   \and
      {PACS-key}{discribing text of that key}
     } 
} 
\maketitle
\section{Introduction}
\label{intro}
Mapping the phase structure of Quantum Chromodynamics (QCD), in particular locating the conjectured QCD critical point (CP), remains a central goal in contemporary nuclear physics~\cite{review}. Lattice QCD calculations at small baryon chemical potential suggest a smooth crossover~\cite{crossover}, while many effective models predict a first-order phase transition at higher baryon density~\cite{1st-PT01,1st-PT02} terminating at a second-order CP~\cite{Stephanov1,Stephanov2}. Experimental confirmation of such a CP would have profound implications for our understanding of strongly interacting matter and the QCD phase transition.

The search for the QCD CP has been a major focus of the RHIC Beam Energy Scan (BES) program~\cite{STAR-note} and will continue at FAIR, NICA, and other heavy-ion facilities. A variety of observables have been proposed to probe the enhanced fluctuations and long-range correlations expected near the CP. These include higher-order cumulants of conserved charges~\cite{Stephanov2,moment02}, factorial moments and intermittency exponents of particle multiplicity distributions~\cite{fac-moment,WuYF-PRL}, and two- and multiparticle correlation functions~\cite{four-part-corr}. These observables are constructed from final-state particle distributions measured within specified momentum and rapidity acceptances, making momentum-space fluctuations an essential part of experimental searches for QCD critical phenomena~\cite{m-STAR-Luo,STAR-WuJ,NA49}.

The study of critical correlations in momentum space also has a well-established theoretical foundation. In renormalization-group theory, the power-law correlations characteristic of critical phenomena can be expressed in both coordinate and Fourier spaces~\cite{ma2000modern}. In the context of QCD critical phenomena, Antoniou et al.~\cite{CMC01,antoniou2016prc} developed an effective-theory description relating critical baryon-density fluctuations in transverse coordinate space to fractal correlations in transverse-momentum space. The resulting momentum-space scaling leads to characteristic intermittency behavior of factorial moments~\cite{CMC01,antoniou2016prc,Antoniou-intermittency,CMC02}, providing a theoretical basis for investigating critical fluctuations through momentum-space particle distributions.

Despite these developments, identifying critical fluctuations in heavy-ion collision data remains challenging. The expected critical contributions may be small compared with conventional non-critical fluctuations~\cite{WuJ-PRC,WangR,WuJ-2021,WuJ-PLB,LiZM,NA49}, while higher-order fluctuation observables often require large event samples~\cite{STAR-note,ChenLZ,Koch-2019}. Moreover, the rapid expansion and finite lifetime of the collision system, together with non-equilibrium evolution and freeze-out effects, can modify the critical correlations that survive in the final state~\cite{non-eq02,non-eq03,non-eq04,non-eq05,LiXB-PRE}. These considerations motivate the development of complementary methods capable of extracting collective correlation structures from event-by-event data.


The eigen-microstate approach (EMA) provides a framework for analyzing such correlations through the spectral decomposition of an event–event correlation matrix~\cite{ChenXS-01,ChenXS-02,ChenXS-03,ChenXS-04,ChenXS-05}. By representing each collision event through its particle multiplicity fluctuations in transverse-momentum space, EMA extracts a hierarchy of orthogonal eigen-microstates and their corresponding weights. Previous studies have shown that characteristic eigen-microstate patterns and changes in the eigenvalue spectrum can emerge in systems containing critical fluctuations~\cite{eigen-1}. These findings suggest that EMA may provide useful information about collective correlations beyond conventional fluctuation observables.

An important step toward experimental applications is to understand how the eigen-microstate patterns and eigenvalue spectra respond to different sources of correlations, particularly when critical fluctuations coexist with conventional heavy-ion collision backgrounds. In this work, we investigate these questions using Ultra-relativistic quantum molecular dynamics (UrQMD) and two stochastic reference ensembles, together with hybrid samples incorporating critical-scaling-motivated correlations generated by the Critical Monte Carlo (CMC) model. By considering both event-level and particle-level incorporation of CMC correlations, we examine the sensitivity of EMA to different signal fractions and investigate the persistence of the resulting eigen-microstate patterns and eigenvalue ratios across momentum-space resolutions.

The paper is organized as follows. Section 2 introduces EMA for relativistic heavy-ion collisions. Section 3 examines the eigen-microstates and eigenvalue spectra of non-critical reference models. Section 4 presents the results for hybrid UrQMD+CMC samples. Section 5 investigates the momentum-space resolution dependence of the EMA signatures. Section 6 summarizes our findings.




\section{The Eigen-Microstate Approach}

In relativistic heavy-ion collisions, the experimentally accessible observables are the phase-space distributions of final-state charged particles. Even under identical macroscopic conditions--collision energy, nuclear species, and impact parameter--the final-state particle distributions fluctuate significantly from event to event. These variations arise from differences in the dynamical evolution of each collision. When a quark-gluon plasma forms, distinct evolution paths correspond to different times of chemical and kinetic freeze-out. Consequently, each collision event is represented by an original microstate (OM), defined through the fluctuations of its final-state particle distribution. The ensemble of such microstates provides the input for the correlation-matrix analysis.

In the application of EMA to relativistic heavy-ion collisions, fluctuations of final-state charged particles in transverse-momentum space are considered here. The $(p_x,p_y)$ plane is discretized into $N = L \times L$ square bins. For the $i$-th event, the corresponding original microstate can be defined by a vector of the multiplicity fluctuations in overall bins of transverse-momentum space,  
\begin{equation}
  x^{i}_k =\Delta N_{{\rm ch},k}^i= N^{i}_{{\rm ch},k} - \langle N_{{\rm ch},k} \rangle, \ \ k=1,...,N; \ i=1,...,M,
\end{equation}
where $N^{i}_{{\rm ch},k}$ is the number of charged particles in the bin $k$ in the event $i$, and $\langle N_{{\rm ch},k} \rangle$ denotes the multiplicity averaged by events in the bin $k$.

We use all charged particles because their substantially larger multiplicity provides better statistical sampling of the transverse-momentum distribution than identified particles. In particular, dividing the momentum plane into $L\times L$ bins leaves identified particles with very sparse occupancy, which can exacerbate finite-statistics effects in the correlation matrix and its eigenstates. Charged particles therefore enable a more reliable investigation of multiscale correlation patterns within the EMA framework.

From the ensemble of $M$ events, we construct the event-event correlation matrix
\begin{equation}
  C_{ij} = \frac{1}{\mathscr{N}} \sum_{k=1}^{N} x^{i}_k\, x^{j}_k,
  \label{eq:Cij}
\end{equation}
which is real and symmetric. $\mathscr{N}=\sum_{i=1}^{M}\sum_{k=1}^{N}(x^{i}_{k})^2$. Diagonalizing $C$ yields eigenvalues $\lambda_n$ and the corresponding linear independent eigenvectors ${\pmb v}^{(n)}$,
\begin{equation}
  \sum_{j=1}^{M} C_{ij} v^{(n)}_j = \lambda_n v^{(n)}_i,
  \qquad n=1,\dots,M.
\end{equation}
The eigenvalues are ordered as $\lambda_1 \geq \lambda_2 \geq \cdots \geq \lambda_M$.

The $n$-th eigen microstate (EM$_n$) is constructed as a linear combination of the original microstates,
\begin{equation}
  \bm{EM}_n = \sum_{i=1}^{M} v^{(n)}_i\, \bm{x}^{i}.
\end{equation}
Its squared norm equals the eigenvalue,
\begin{equation}
  |\bm{EM}_n|^2 =  \lambda_n.
\end{equation}
The eigenvalues are non-negative and satisfy $\sum_{n=1}^{M}\lambda_n=1$ due to the normalization of the correlation matrix. Therefore, the eigenvalues represent the normalized weights of the corresponding eigen-microstates, $w_n=\lambda_n$.

In particular, the largest eigenvalue $w_1=\lambda_1$ measures the relative contribution of the dominant fluctuation mode to the total fluctuation strength. A larger $w_1$ indicates that a greater fraction of the fluctuations is concentrated in the leading eigen-microstate, while the remaining eigenvalues characterize the contributions from subleading modes.

The eigenvalue spectrum and the corresponding eigen-microstate patterns provide complementary information about the collective correlation structure of the event ensemble. Changes in the leading weights, together with the emergence of coherent multiscale patterns, can thus be used to characterize the response of EMA to the correlations introduced by the CMC model.

Note that each eigenvector is defined only up to an overall sign: flipping the sign of an EM leaves the equation and the eigenvalue $\lambda_n$ unchanged. Therefore, the overall sign of an EM has no physical meaning. The physically relevant information lies in the spatial shape, localization, and relative amplitude of the EMs and in the hierarchy of their weights.

The cumulative weight is defined as
\begin{equation}
  C(m) = \sum_{n=1}^{m} w_n,
\end{equation}
where $m$ is the number of eigenvalues. $C(m)$ characterizes how the total fluctuation strength is distributed among the leading eigen-microstates. A more rapid saturation indicates that a smaller number of leading modes accounts for a larger fraction of the total spectral weight.

To characterize the relative contributions of different fluctuation modes, we further consider the ratios of subleading to leading eigenvalue weights $\frac{w_n(L,\alpha)}{w_1(L,\alpha)}$, with $n=2,3$, where $L$ denotes the number of bins along each direction of the transverse-momentum plane, and $\alpha$ represents the fraction of the CMC component introduced into the event ensemble.

These dimensionless ratios characterize the relative spectral weights of the corresponding eigen-microstates. By examining their dependence on $L$, we investigate whether the correlation structures extracted by EMA remain stable under changes in momentum-space resolution.

\section{Eigen Microstates of Non-Critical Models}

To understand how conventional dynamical correlations influence the eigenvalue spectra and EM patterns, we first study non-critical reference systems. We consider UrQMD, and two stochastic models that share the same event-by-event multiplicity distribution as that of UrQMD, but differ in whether they impose global kinetic constraints. 

As a realistic non-critical baseline, UrQMD~\cite{urqmd01,urqmd02} incorporates essential non-critical dynamics of heavy-ion collisions, including energy--momentum and charge conservation, resonance production and decay, hadronic rescatterings, and collective flow. We will examine to what extent these conventional mechanisms leave an imprint on the EM patterns. 

We generate 0--5\% most-central Au+Au collisions at $\sqrt{s_{\rm NN}} = 19.6$~GeV using UrQMD (v3.4). Final-state charged particles are selected within the STAR acceptance of high moment analysis~\cite{m-STAR-Luo,STAR-WuJ}: pseudorapidity $|\eta| < 0.5$, transverse momentum windows $0.2 < p_{\rm t} < 1.6~\mathrm{GeV}/c$ for $\pi^{\pm}$ and $K^{\pm}$, $0.4 < p_{\rm t} < 2.0~\mathrm{GeV}/c$ for $p$ and $\bar{p}$. These cuts are applied to all models and hybrid samples analyzed in this work. 

To separate the influence of global spectral shapes from that of true dynamical correlations, we construct two stochastic reference ensembles:

\begin{itemize}
  \item \textbf{Stochastic model I without kinetic constraints.}  
  In each event, the total charged multiplicity is drawn from the same multiplicity distribution as UrQMD. The particle momenta are then assigned independently and uniformly in $p_x,p_y \in (-1,1)~\mathrm{GeV}/c$. No azimuthal modulation or transverse-momentum structure is imposed; this ensemble contains no inter-particle correlations and no common momentum shape.
  \item \textbf{Stochastic model II with kinetic constraints.}  
  Events again share the UrQMD multiplicity distribution, but transverse momenta of the particle $p_{\rm t}$ are drawn from the global momentum distribution $
p(p_{\rm t}) = \frac{1}{\Gamma(k)\,\beta^{k}}\, p_{\rm t}^{k-1}\, {\rm e}^{-p_{\rm t}/\beta}$ where $\Gamma(k)$ is the Gamma function, $k$ and $\beta$ are parameters which are obtained by fitting the $p_{\rm t}$ spectrum of UrQMD samples. The azimuthal angle $\phi$ is sampled from a weakly anisotropic flow-modulated distribution
$p(\phi) = \frac{1}{2\pi}\left(1 + 0.08 \cos 2\phi\right)$. These constraints impose a global momentum-shape structure (a realistic $p_{\rm t}$ spectrum and small elliptic-flow-like modulation) but do not introduce any intrinsic particle--particle correlations.
\end{itemize}

These two stochastic ensembles thus provide clean baselines ``without'' and ``with'' kinetic constraints, allowing us to isolate the role of global momentum shapes from that of UrQMD-specific dynamical correlations.

\begin{figure}[t]
\resizebox{0.5\textwidth}{!}{%
  \includegraphics{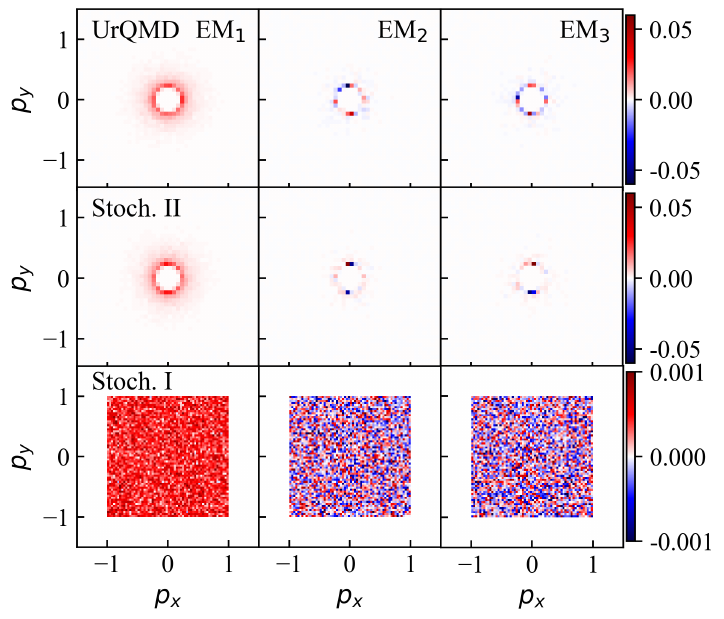}
}
\caption{(Color online) First three EMs (EM$_1$, EM$_2$, EM$_3$) for UrQMD and for the two stochastic non-critical ensembles: Stoch. II and Stoch. I with and without kinetic constraints at $L=60$.}
\label{fig:em_noncritical}
\end{figure}

Figure~\ref{fig:em_noncritical} displays the first three EMs (EM$_1$, EM$_2$, EM$_3$) for UrQMD and the two stochastic reference models. Each EM is shown as a color map in the $(p_x,p_y)$ plane. Red and blue denote positive and negative values of $\Delta N_{\rm ch}$. Several non-critical features are evident:

\begin{itemize}
  \item EM$_1$ is essentially \emph{identically colored} for all three ensembles. Because the overall sign of an eigenvector is arbitrary, EM$_1$ appears identically red. This corresponds to an average global mode that reflects the overall radial momentum distribution with no internal cluster structure.
  \item EM$_2$ and EM$_3$ display \emph{random red--blue fluctuations} with no stable patch-like or directional structure. They resemble noise patterns superimposed on the broad average global mode.
\end{itemize}

%
%

%

These observations indicate that the conventional correlations represented by the investigated non-critical models do not generate the coherent cluster-like EM patterns observed in the CMC-containing samples. Within these reference ensembles, EMA distinguishes the modeled multiscale correlations from the non-critical background structures.

\begin{figure}[t]
\centering\resizebox{0.35\textwidth}{!}{
  \includegraphics{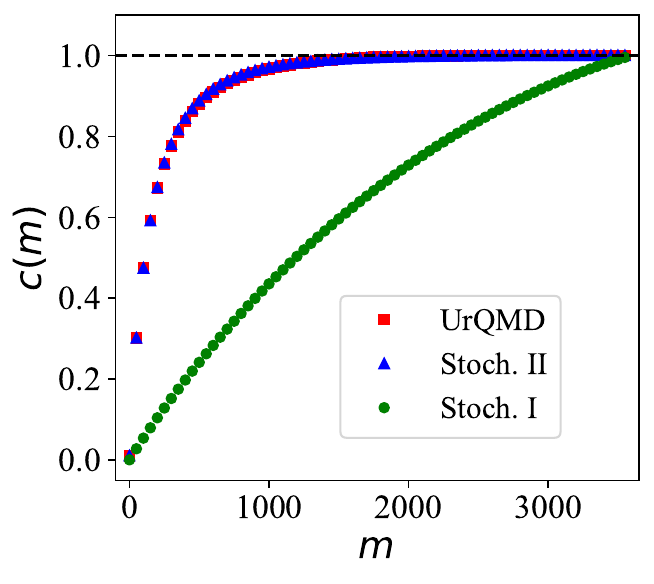}}
  \caption{(Color online) Cumulative weight $C(m)$ for the UrQMD ensemble, the stochastic model II and I with and without kinetic constraints at $L=60$.}
  \label{fig:cumulant_noncritical}
\end{figure}

To quantify the eigenvalue hierarchy, Fig.~\ref{fig:cumulant_noncritical} shows the cumulative weight $C(m)$ for the three ensembles. In the uniform stochastic model I (without kinetic constraints), $C(m)$ increases almost linearly with $m$, indicating that there is no preferred mode and the weights are distributed broadly over many EMs. By contrast, both the kinetic-constraint stochastic model II and UrQMD exhibit a rapid initial rise of $C(m)$ followed by early saturation, reflecting the presence of a common global momentum shape that enhances a small number of leading non-critical modes.

The close agreement between UrQMD and stochastic model II suggests that, for the investigated ensembles, the leading EMA structures are primarily governed by the global transverse-momentum and azimuthal distributions, with relatively small contributions from the additional dynamical correlations included in UrQMD.

These results provide a practical baseline for future experimental applications of EMA. Such spectral and pattern features may serve as useful reference characteristics of non-critical backgrounds in future experimental applications of 
EMA.

\section{Eigen Microstates of Hybrid Models}

We now introduce critical fluctuations into the UrQMD baseline using a CMC model and investigate how the EM patterns and eigenvalue spectra respond. In relativistic heavy-ion collisions, whether a collision trajectory passes through the critical region depends on the collision conditions, such as beam energy and collision centrality. Consequently, critical fluctuations may appear in different forms. One possible scenario is that only a fraction of events passes through the critical region, while the remaining events evolve entirely outside it. Another possibility is that only localized domains within an event experience critical dynamics, for example within localized regions of the evolving QCD matter, so that only a subset of particles carries critical correlations whereas the remaining particles exhibit only conventional short-range correlations. 

These two limiting scenarios are approximately modeled here by event-level and particle-level embedding of CMC events into the UrQMD background, respectively. The purpose is to investigate how the EM patterns evolve with increasing critical fraction under these two representative situations. These two constructions provide complementary settings for investigating the EMA response to critical correlations introduced at different levels of the event ensemble.

\subsection{Critical Monte Carlo model}

The CMC model~\cite{CMC01,antoniou2016prc,CMC02,Antoniou-intermittency} is constructed to generate scale-invariant fluctuations in transverse-momentum space based on the critical properties of the three-dimensional Ising universality class, to which the QCD critical point is expected to belong.

As demonstrated by Antoniou et al.~\cite{CMC01,antoniou2016prc}, the power-law correlations of baryon-density fluctuations in transverse coordinate space can be transformed into corresponding power-law correlations in transverse-momentum space through a Fourier transformation. The resulting momentum-space fractal dimension is directly related to the critical isotherm exponent. These correlations lead to characteristic intermittency scaling of factorial moments with the number of momentum-space bins~\cite{CMC02,Antoniou-intermittency,WuJ-PLB,WuJ-PRC,WangR,WuJ-2021,LiZM}.

The CMC model employs Lévy random walks to generate particle momentum distributions with fractal correlations characterized by the theoretically predicted critical scaling exponents. The corresponding intermittency behavior has been investigated in previous theoretical and phenomenological studies~\cite{CMC02,Antoniou-intermittency,WuJ-PRC,WangR,WuJ-2021,WuJ-PLB,LiZM}. These studies establish CMC as an effective realization of the momentum-space scaling properties motivated by QCD critical theory.

In the present study, CMC is used to construct final-state momentum distributions containing prescribed multiscale correlations. By embedding these distributions into the UrQMD background, we investigate the response of the eigen-microstate patterns and eigenvalue spectra to different fractions of critical-scaling-motivated correlations. This effective construction focuses on the resulting momentum-space correlation structures, while the dynamical formation and evolution of critical fluctuations during the collision are left for future investigations.

\subsection{Construction of hybrid UrQMD+CMC samples}

To investigate how EMA responds to critical fluctuations embedded in a realistic heavy-ion background, we construct hybrid event ensembles that combine UrQMD and CMC. We consider two physically distinct scenarios:

\begin{itemize}
  \item \textbf{Replacement of event-level (fraction $\alpha_{\rm e}$).}  
  A fraction $\alpha_{\rm e}$ of events in the UrQMD ensemble is completely replaced by CMC events. In this case, some events are purely non-critical (UrQMD) while others are fully critical (CMC). This scenario mimics a situation in which only a subset of heavy-ion collisions traverse the critical region in the QCD phase diagram.
  \item \textbf{Particle-level replacement (fraction $\alpha_{\rm p}$).} 
 In each UrQMD event, a fraction $\alpha_p$ of final-state particles is replaced by particles drawn from a CMC event. This construction is motivated by the possibility that critical fluctuations develop within localized regions of the collision system, so that only a subset of the final-state particles carries critical correlations. The replacement introduces these correlations directly into the final-state momentum distribution, providing an effective representation of such a scenario. To avoid distorting the UrQMD $p_{\rm t}$ spectrum, we impose a $p_{\rm t}$-matching condition on each replacement:
  \begin{equation}
    \bigl| p_{\rm t}^{\mathrm{CMC}} - p_{\rm t}^{\mathrm{UrQMD}} \bigr| < 0.2~\mathrm{GeV}/c.
  \end{equation}
\end{itemize}

\subsection{Emergence of critical cluster patterns in eigen microstates}

\begin{figure*}
\resizebox{1.0\textwidth}{!}{
  \includegraphics{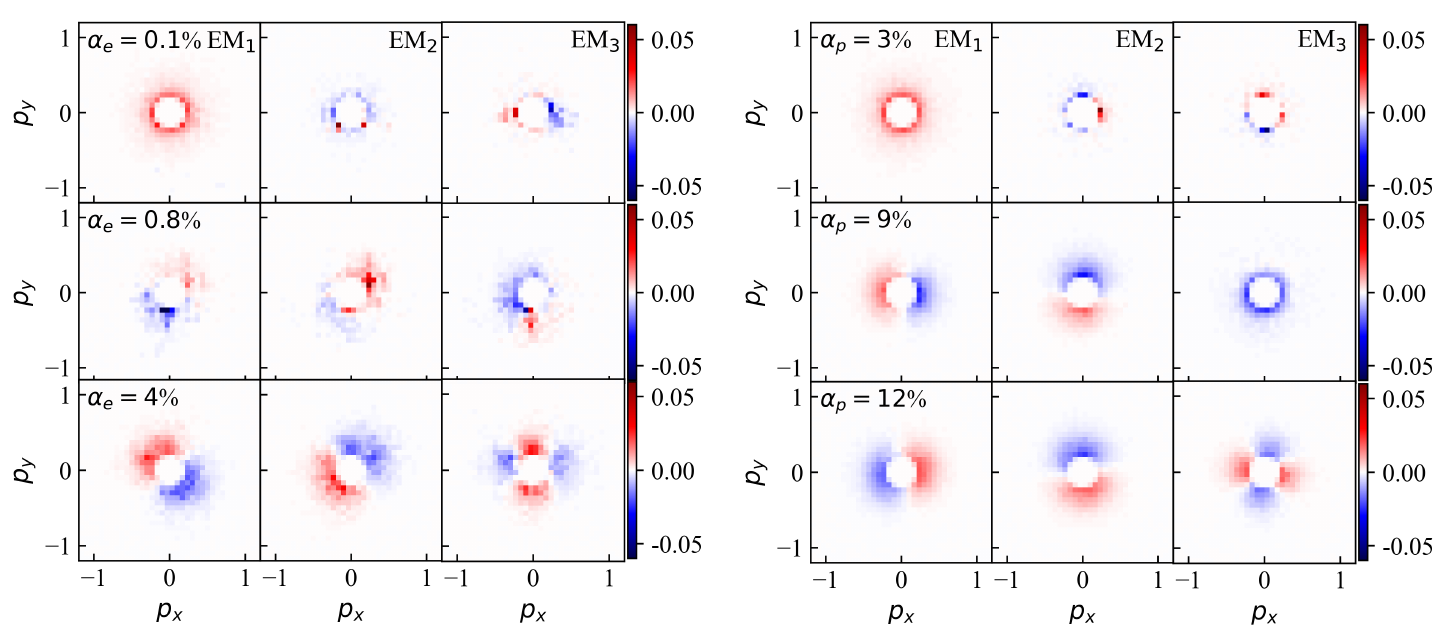}}
  \caption{(Color online) First three EMs for hybrid UrQMD+CMC ensembles constructed with event-level replacement (left columns) and particle-level replacement (right columns) for several critical fractions $\alpha_{\rm e}$ and $\alpha_{\rm p}$ at $L=60$. }
  \label{fig:em_hybrid}
\end{figure*}

Figure~\ref{fig:em_hybrid} shows the first three EMs for hybrid ensembles constructed with event-level replacement (left columns) and particle-level replacement (right columns) in several critical fractions.

For \textbf{event-level replacement}, we observe:

\begin{itemize}
  \item At $\alpha_{\rm e} = 0.1\%$, EM$_1$ still resembles the UrQMD baseline pattern, while EM$_2$ and EM$_3$ begin to exhibit a weak separation of the red and blue regions, signaling the onset of a new collective mode associated with criticality.
  \item At $\alpha_{\rm e} = 0.8\%$, clear patch- or cluster-like structures appear in all three leading EMs. These patterns break the approximate rotational symmetry seen in the non-critical baseline and exhibit localized regions of enhanced positive or negative fluctuation.
  \item At $\alpha_{\rm e} = 4\%$, the cluster-like patterns in EM$_1$--EM$_3$ become strong and stable: EM$_1$ is dominated by a coherent two-patch structure, while EM$_2$ and EM$_3$ show subleading but still well-organized patch-patterns.
\end{itemize}

For \textbf{particle-level replacement}, the emergence of cluster-like EM patterns is qualitatively similar but requires significantly larger critical fractions $\alpha_{\rm p}$:

\begin{itemize}
  \item At $\alpha_{\rm p} = 3\%$, EM$_1$ remains close to the baseline pattern and EM$_2$, EM$_3$ appear in the embryonic form of a cluster-like pattern.
  \item At $\alpha_{\rm p} = 9\%$, the development of cluster-like patterns is more visible in EM$_1$ and EM$_2$. EM$_3$ continues to be a baseline pattern; however, its weight gradually decreases, exhibiting a vanishing trend.
  \item At $\alpha_{\rm p} = 12\%$, all three leading EMs exhibit clearly developed cluster-like patterns. 
\end{itemize}

The comparison between event-level and particle-level replacement demonstrates that EMA is particularly sensitive to \emph{coherent event-level criticality}: a very small fraction of fully critical events ($\alpha_{\rm e} \sim 0.1\%$--$1\%$) already leave a visible imprint on the EM patterns. By contrast, localized critical droplets require substantially larger particle-level fractions ($\alpha_{\rm p} \gtrsim 9\%$) to produce comparable cluster structures. This difference reflects the fact that EMs are collective modes of the entire event ensemble and thus respond more strongly to coherent event-level structure than to local perturbations. 

These results provide important guidance for experimental searches: if criticality is realized only in a small subset of events at certain collision energies or centralities, its earliest signatures in EMA are expected to appear as the emergence and subsequent sharpening of cluster-like structures from a UrQMD background. 

In our hybrid construction and stochastic baselines, exact energy-momentum conservation is not strictly enforced. However, the absence of any cluster-like patterns in both stochastic models (without momentum conservation) and the UrQMD (with momentum conservation) indicates that conservation-law violations by themselves do not appear to generate the observed coherent EM structures. Instead, the emergence of such structures is associated with the presence of long-range, scale-invariant correlations introduced by the CMC component.

Near a CP, the correlation length diverges in the thermodynamic limit. In finite systems, this manifests as correlations spanning a broad range of length scales. In our hybrid model, such critical correlations are introduced via Lévy walks in transverse momentum space, effectively mimicking scale-invariant fluctuations. The leading eigen-microstates provide orthogonal representations of the dominant correlation structures, with their spatial patterns reflecting different characteristic features of the underlying multiscale fluctuations. The leading eigen-microstates exhibit different spatial structures, ranging from broad two-patch patterns to more intricate multi-patch configurations. These patterns provide complementary representations of the multiscale correlations introduced by CMC.
The observed multi-patch structures are consistent with the multiscale correlations implemented in the CMC model.

\subsection{Eigenvalue weights and their cumulative distributions}

\begin{figure}[t]
  \resizebox{0.5\textwidth}{!}{
  \includegraphics{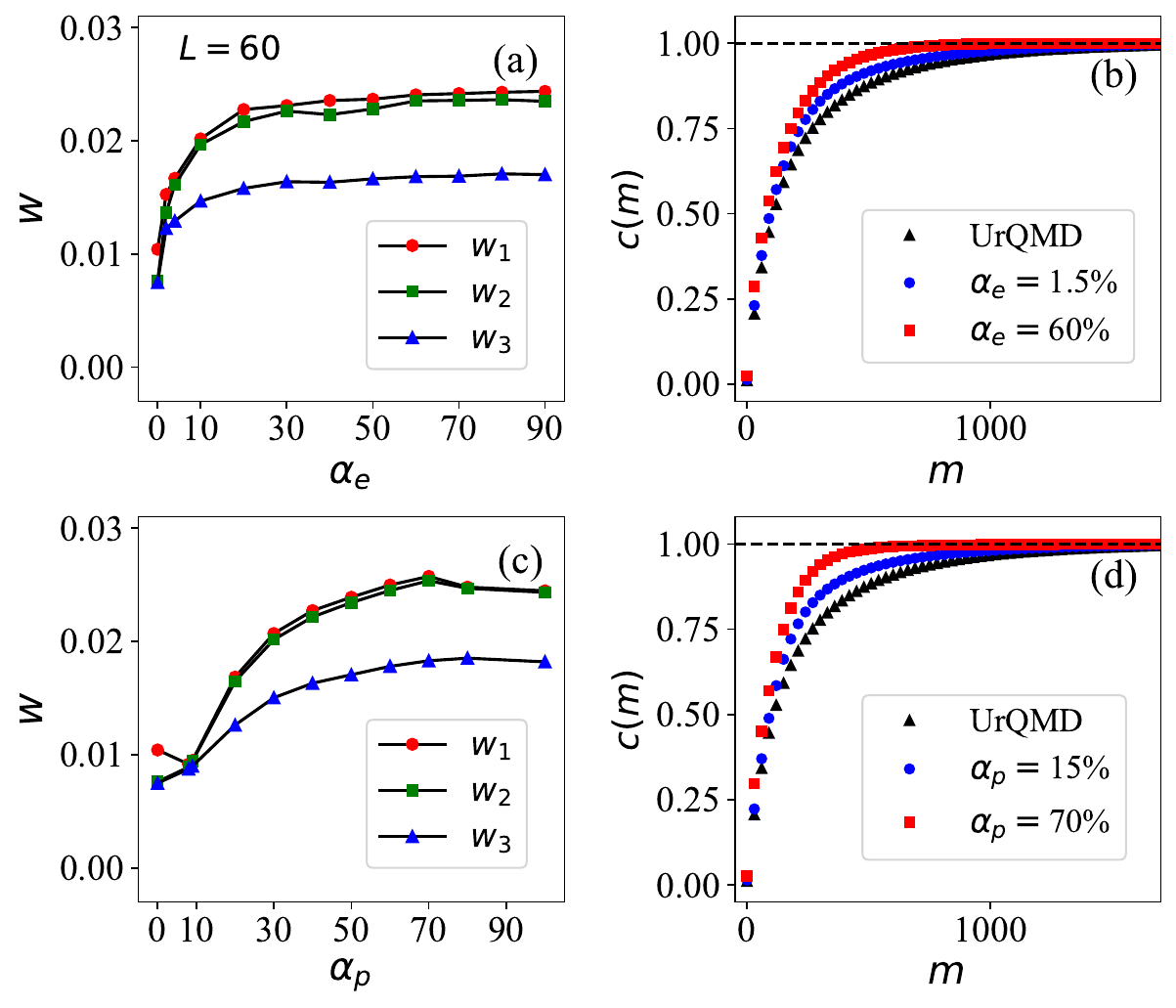}}
  \caption{(Color online) Top three eigenvalues $w_1$, $w_2$, $w_3$ as functions of the critical fractions $\alpha_{\rm e}$ (event-level replacement) and $\alpha_{\rm p}$ (particle-level replacement), together with the corresponding cumulative weight $C(m)$ at $L=60$. }
  \label{fig:eig_hybrid}
\end{figure}

Complementary information is provided by the eigenvalue hierarchy. 
Figure~\ref{fig:eig_hybrid} shows the three largest weights $w_1$, $w_2$, $w_3$ as a function of $\alpha_{\rm e}$ (event-level replacement) and $\alpha_{\rm p}$ (particle-level replacement), together with the corresponding cumulative weight $C(m)$.

For event-level replacement [Fig.~\ref{fig:eig_hybrid}(a)], the two leading weights $w_1$ and $w_2$ increase with $\alpha_{\rm e}$ and approach similar saturated values, while $w_3$ remains smaller. This behavior indicates that the CMC-induced correlations become increasingly concentrated in a few leading eigen-microstates. Consistently, the cumulative weight $C(m)$ in Fig.~\ref{fig:eig_hybrid}(b) rises more rapidly and approaches saturation at smaller $m$ as the CMC fraction increases.

For particle-level replacement [Fig.~\ref{fig:eig_hybrid}(c)], $w_1$ displays a more complex behavior. At very small $\alpha_{\rm p}$, it can exhibit a slight decrease, reflecting the disruption of the baseline mode by a small number of critical signal particles. Once the critical fraction becomes sufficiently large, $w_1$ begins to increase and eventually saturates at large $\alpha_{\rm p}$. The corresponding cumulative weights $C(m)$ in Fig.~\ref{fig:eig_hybrid}(d) show an analogous trend: with increasing $\alpha_{\rm p}$ the rise of $C(m)$ becomes steeper and saturation occurs at smaller $m$, indicating that a few leading EMs increasingly dominate the ensemble.

Taken together, these results demonstrate how the introduction of CMC correlations modifies the eigenvalue spectrum of the event ensemble. The leading weight $w_1$ measures the relative contribution of the dominant fluctuation mode, while the cumulative weight $C(m)$ characterizes the distribution of fluctuation strength among the eigen-microstates.

As the CMC contribution increases, the spectral weights become increasingly concentrated in a small number of leading modes, accompanied by the development of coherent cluster-like eigen-microstate patterns. The different responses observed for event-level and particle-level replacement reflect the different ways in which the CMC-induced correlations are incorporated into the event ensemble.

The present analysis is performed with event ensembles of size $M=20,000$, which is sufficient to reveal the qualitative hierarchy of eigenvalues and EM patterns. We have examined a larger ensemble of events and found that the main features, including the emergence of cluster-like structures and the hierarchy of leading eigenvalues, remain stable. 

\section{Momentum-Space Resolution Dependence of EMA Signatures}

Scale-invariant correlations are an important characteristic of critical phenomena. As discussed in Section 4.1, the CMC model generates transverse-momentum correlations with fractal scaling properties. To investigate whether EMA can capture these multiscale correlations, we examine the dependence of the eigen-microstate patterns and eigenvalue ratios on the momentum-space resolution $L$, defined as the number of bins along each direction of the fixed transverse-momentum region.  

\begin{figure}[t]
  \resizebox{0.5\textwidth}{!}{
  \includegraphics{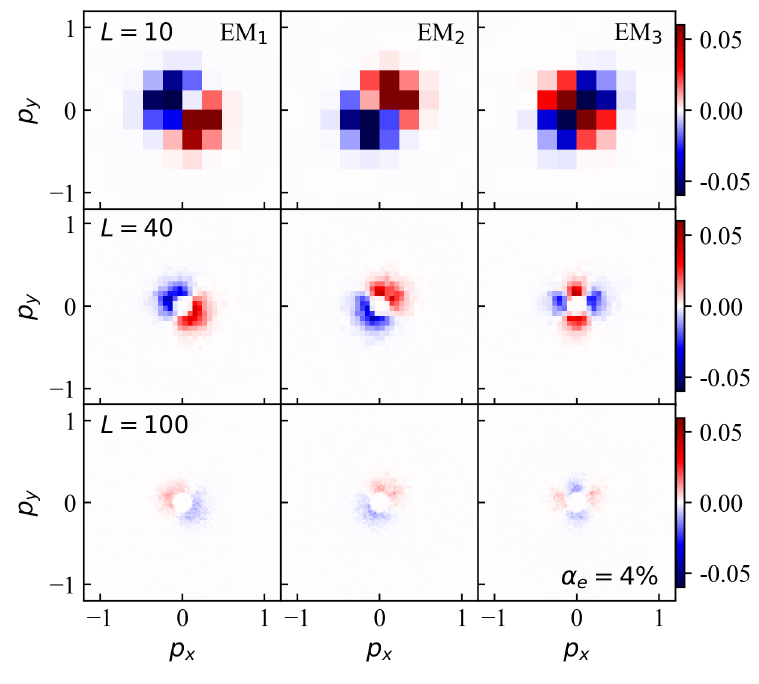}}
  \caption{(Color online) First three EMs at fixed $\alpha_{\rm e} = 4\%$ for three momentum-space resolutions $L=10$, $40$, and $100$.}
  \label{fig:em_fss}
\end{figure}

Figure 5 presents the first three eigen-microstates for hybrid ensembles with $\alpha_{\rm e}=4\%$ at three resolutions, $L=10$, 40, and 100. As $L$ increases, the patterns become more dilute because of the finer binning. Nevertheless, the underlying patch-like structures remain clearly visible and morphologically similar. In particular, EM$_1$ and EM$_2$ exhibit two dominant patches, whereas EM$_3$ displays a more complex multi-patch structure. The persistence of these characteristic patterns across different resolutions indicates that the eigen-microstates retain important features of the multiscale momentum-space correlations introduced by CMC.

To quantify the resolution dependence of the eigenvalue spectrum, we examine the ratio $w_3/w_1$ as a function of the CMC fraction for $L=40$, 60, and 100, as shown in Fig. 6. At small CMC fractions, the ratios differ among the three resolutions, reflecting the combined effects of the non-critical background and resolution-dependent sampling. As the CMC contribution increases, the differences gradually decrease. For event-level replacement, the ratios become approximately resolution-independent at $\alpha_{\rm e}\gtrsim4\%$, while a similar behavior is observed for particle-level replacement at sufficiently large $\alpha_{\rm p}$.

\begin{figure}[t]
  \resizebox{0.5\textwidth}{!}{
  \includegraphics{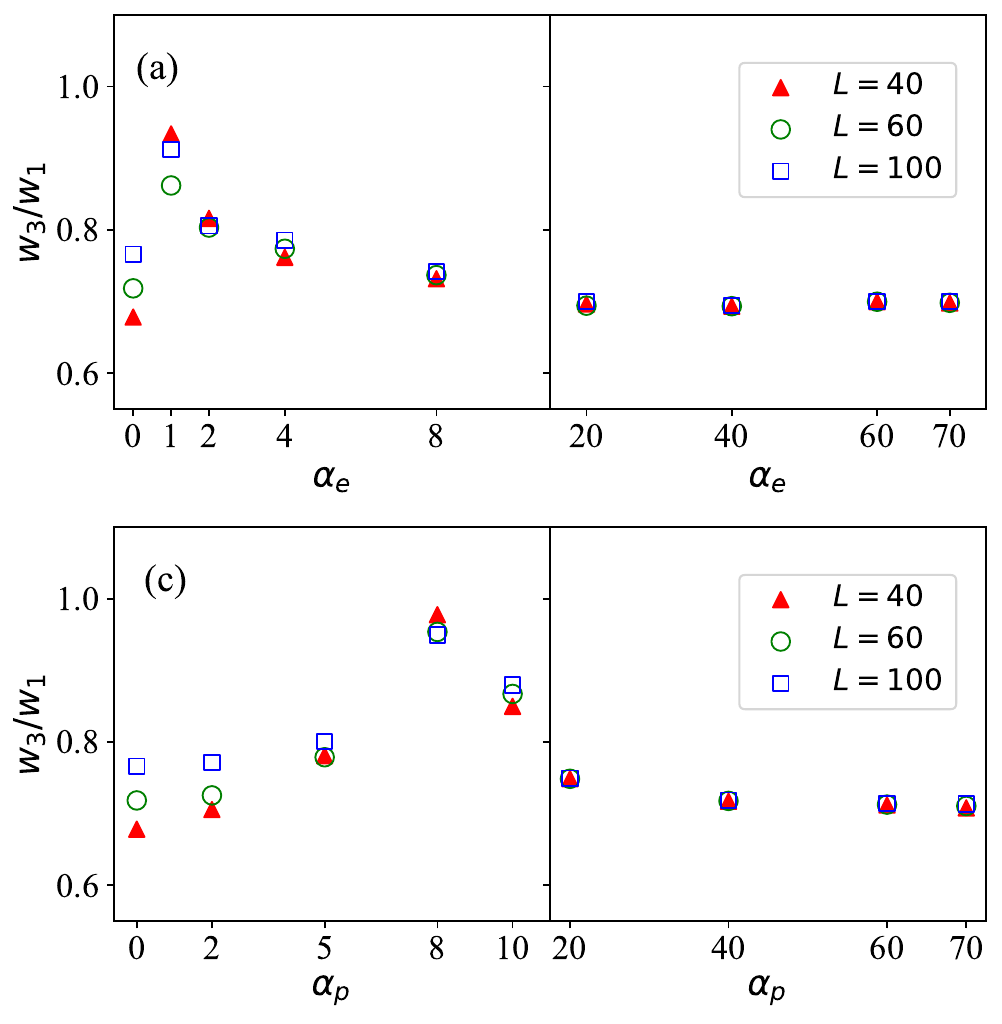}}
  \caption{(Color online) Ratio $w_3/w_1$ as a function of $\alpha_{\rm e}$ and $\alpha_{\rm p}$ for momentum-space resolutions $L=40$, $60$, and $100$. }
  \label{fig:ratio_fss}
\end{figure}

The approximate convergence of $w_3/w_1$ across different momentum-space resolutions indicates that the relative spectral contributions of the leading eigen-microstates become increasingly stable as the CMC-induced correlations become more prominent. Together with the persistence of the characteristic cluster-like patterns shown in Fig. 5, this behavior demonstrates that EMA captures important multiscale features of the momentum-space correlations generated by CMC. 

\section{Summary and Conclusions}

In this work, we have performed a systematic model study of the EMA for characterizing critical-scaling-motivated correlations. By analyzing UrQMD, two stochastic reference ensembles, and hybrid UrQMD+CMC samples, we have investigated the response of EMA to conventional non-critical correlations and to multiscale momentum-space correlations introduced by the CMC model.

For the non-critical reference ensembles, we find that the leading eigen-microstates and eigenvalue spectra are primarily influenced by the global transverse-momentum and azimuthal distributions. The leading eigen-microstate exhibits a smooth global structure, whereas the subleading eigen-microstates show no coherent cluster-like patterns. The close agreement between UrQMD and the stochastic model with similar global momentum distributions indicates that the additional dynamical correlations included in UrQMD have a relatively small effect on the leading eigen-microstate structures under the conditions investigated.

We then introduced CMC correlations into the UrQMD background through two simplified scenarios: event-level and particle-level replacement. In the event-level scenario, even a small fraction of CMC events produces visible changes in the subleading eigen-microstates, and coherent cluster-like patterns emerge as the CMC fraction increases. In contrast, particle-level replacement requires substantially larger CMC fractions to produce comparable structures. This difference demonstrates that EMA responds more strongly to correlations introduced coherently at the event level than to those embedded within individual events.

The introduction of CMC correlations also modifies the eigenvalue spectrum. The largest eigenvalue $w_1$ measures the relative contribution of the dominant fluctuation mode to the total fluctuation strength, while the cumulative weight $C(m)$ characterizes the concentration of spectral weight among the leading eigen-microstates. As the CMC fraction increases, the leading modes become more prominent and the cumulative weight saturates more rapidly. Furthermore, the characteristic cluster-like patterns persist across different momentum-space resolutions, while the eigenvalue ratio $w_3/w_1$ exhibits approximately resolution-independent behavior when the CMC contribution becomes sufficiently strong. These observations indicate that EMA captures important multiscale features of the momentum-space correlations generated by CMC.

The present results demonstrate the potential of EMA as a complementary method for characterizing critical-scaling-motivated correlations in the presence of non-critical heavy-ion collision backgrounds. The CMC model provides an effective realization of momentum-space correlations motivated by QCD critical theory, instead of a dynamical description of their formation and evolution. Consequently, quantitative applications to experimental data will require further consideration of critical dynamics, freeze-out effects, and particle-species-dependent correlations. Within these limitations, the present study provides a useful basis for applying EMA to future searches for the QCD critical point.

\begin{acknowledgement}
 We acknowledge valuable discussions with Xiaosong Chen, Yufu Lin and Lizhu Chen. This research was funded by the National Key Research and Development Program of China, Grant No.
2024YFA1610700, and 2022YFA1604900, the
National Natural Science Foundation of China, Grant
No. 12275102, and the Guangxi Natural Science Foundation, Grant No. 2025\-GXNSFBA\-069070.
\end{acknowledgement}

%
%

%
 \bibliographystyle{elsarticle-num}
 \bibliography{ref-3}
%

\end{document}